\documentclass[aip]{revtex4-1}
\usepackage{dcolumn}%
\usepackage{bm}%
\usepackage{epsfig}
\usepackage{latexsym}
\usepackage{epstopdf}

\usepackage{amsmath,amssymb}
\usepackage{graphicx}

\begin{document}
\title{Wigner crystals for a planar, equimolar binary mixture of classical, 
charged particles%
%\thanks{``\ldots{}I wouldn't stand
%by and see the rules broken -- because right is right, and wrong
%is wrong, and a  body ain't got no business doing wrong when he
%ain't ignorant and knows better.'' \protect\\ [0.9ex] \strut\qquad
%Twain~M., The Adventures of Huckleberry Finn. 1884.}
}
\author{Moritz Antlanger}
\email[]{moritz.antlanger@tuwien.ac.at}
%\homepage[]{Your web page}
%\thanks{}
\affiliation{Institut f\"ur Theoretische Physik and Center for Computational Material Science (CMS), Vienna University of Technology, AUSTRIA} 
\affiliation{Laboratoire de Physique Th\'eorique (UMR 8627), Universit\'e de Paris-Sud and CNRS, B\^atiment 210, 91405 Orsay Cedex, FRANCE } 
\author{Gerhard Kahl}
%\email[]{Your e-mail address}
%\homepage[]{Your web page}
%\thanks{}
%\altaffiliation{}
\affiliation{Institut f\"ur Theoretische Physik and Center for Computational Material Science (CMS), Vienna University of Technology, AUSTRIA}
%
%% or for single author or if all authors are from the same institute:
%
%  \author[Short authors list]{1st Author, 2st Author, \ldots}
%  \address{Institute}
%
%% Fields in square brakets (short title and short authors list) are
%% optional. Use them if your entries exceeds 45 characters.
%

%\linespread{2.}

\begin{abstract}
  We have investigated the ground state configurations of an
  equimolar, binary mixture of classical charged particles (with
  nominal charges $Q_1$ and $Q_2$) that condensate on a neutralizing
  plane. Using efficient Ewald summation techniques for the
  calculation of the ground state energies, we have identified the
  energetically most favourable ordered particle arrangements with the
  help of a highly reliable optimization tool based on ideas of
  evolutionary algorithms. Over a large range of charge ratios, $q =
  Q_2 / Q_1$, we identify six non-trivial ground states, some of which
  show a remarkable and unexpected structural complexity. For $0.59
  \lesssim q < 1$ the system undergoes a phase separation where the
  two charge species populate in a hexagonal arrangement spatially
  separated areas.
\end{abstract}

\maketitle

\section{Introduction}
\label{sec:introduction}

The identification of the ordered ground state configurations of
classical charged particles is known in literature as the Wigner
problem \cite{ref:wigner_1934}. In two dimensions and at vanishing
temperature these charges form a hexagonal lattice
\cite{ref:meissner_1976, ref:bonsall_1977, ref:levin_2002}. In rather
recent investigations, this problem has been extended: one example is
the bilayer problem where the charges are confined between two
planes, which are separated by a finite distance \cite{ref:samaj_2012, ref:samaj_2012_a}. Due to
the availability of closed, analytic expressions for the potential
energy of this particular system, the complete set of its ground state
configurations could be identified.

In the present contribution we return to the single layer problem and
consider an equimolar, binary mixture of charged particles (with
nominal values $Q_1$ and $Q_2$), that condensate at vanishing
temperature on a neutralizing plane. For given values $Q_1$ and $Q_2$
the ordered equilibrium configurations of the particles at vanishing
temperature are imposed by the requirement that the potential energy
is minimized. With the help of efficient and highly accurate Ewald
summation techniques \cite{ref:mazars_2011}, the lattice sum of this
system can be evaluated for any particle arrangement on an arbitrary,
two-dimensional lattice. Employing suitable optimization
techniques, the parameters of these lattices are then optimized in such
a way as to minimize the lattice sum. In this contribution we have
used an optimization tool that is based on ideas of evolutionary
algorithms (EAs) \cite{ref:holland_1975, ref:gottwald_2005}. Within
this concept, any possible two-dimensional lattice is considered as an
individual, to which a fitness value is assigned. These individuals
are then exposed on the computer to an artificial evolution: via
creation and mutation operations a large number of individuals is
produced; in the former procedure a pair of new individuals is created
from a pair of parent individuals which are selected according to
their fitness values. Along this evolution only the best, {\it i.e.},
the fittest, individuals are expected to survive and are thus
retained. Bearing in mind that we are looking for the individual (=
structure) with the lowest lattice sum, we assign a high fitness value
to an energetically favourable ordered structure.  EA-based
optimization algorithms have turned out to be highly efficient and
reliable tools for identifying ordered equilibrium structures in a
broad variety of condensed matter systems, in general
\cite{ref:gottwald_2005, ref:pauschenwein_2008, ref:kahn_2009,
  ref:kahn_2010}, and for quite a few two-dimensional systems, in
particular \cite{ref:fornleitner_2008, ref:fornleitner_2008_a,
  ref:fornleitner_2009, ref:doppelbauer_2010, ref:antlanger_2011}.

 In total we have identified six non-trivial ground states. They
  are characterized by a broad variety of structural complexity, which
  is the result of an intricate competition between the interactions
  of the two charges. Introducing the charge ratio $q$ as the only
  relevant parameter of the system (defined as $q = Q_2/Q_1$ with $0
  \le q \le 1$ due to simple consideration and due to symmetry
  arguments) we identified two structures that show a remarkable
  stability over relatively large $q$-ranges: (i) one of them can be
  described via two intertwining, commensurate square sublattices, one
  of them populated by charges $Q_1$, the other by charges $Q_2$; this
  structure shows, in addition, among all ground states the highest
  energy gain with respect to a suitably defined reference state; (ii)
  in the other ordered equilibrium configuration strongly distorted,
  but symmetric hexagonal tiles cover the entire two-dimensional space
  hosting in their interiors pairs of the weaker charges. For $0.59
  \lesssim q < 1$ the system undergoes a phase separation where the two
  spatially separated phases are represented by hexagonal lattices
  populated by either species of charges. The remaining four
  non-trivial ground states are dominated by distorted, asymmetric
  hexagonal arrangements of charges $Q_1$, hosting in their interior
  pairs and triplets of charges $Q_2$. For the limiting values, {\it
    i.e.}, $q = 0$ and $q = 1$, we obtain the expected hexagonal
  particle arrangements.

The paper is organized as follows. In the subsequent section we
briefly introduce our model system. Section 3 is dedicated to the
methods we have used: both the Ewald summation technique as well as
our optimization tool (based on ideas of evolutionary algorithms) are
briefly summarized; further we introduce a suitable state of reference
for our energetic considerations. In section 4 we thoroughly discuss
the results. The contribution is closed with concluding remarks.

\section{Model}
\label{sec:model}

We consider an equimolar mixture of classical charges with the
particles being confined to a planar ({\it i.e.}, two-dimensional )
geometry. The point charges (with nominal values $Q_1$ and $Q_2$) are
located at positions ${\bf r}_i$ and ${\bf r}_j$ and interact via an
unscreened Coulomb interaction

\begin{equation}
\Phi(r_{ij}) = \frac{Q_i Q_j}{r_{ij}}
\label{eq:coulomb}
\end{equation}
with $r_{ij} = |{\bf r}_i - {\bf r}_j |$.

Since the total number density ({\it i.e.}, number of particles per
unit area), $\rho$, can be scaled out via the distances, its actual
quantity is irrelevant for further considerations. In the equimolar case we obtain for the partial number
densities $\rho_1 = \rho_2 = \rho/2$.

For convenience we introduce the parameter $q = Q_2 / Q_1$, {\it
  i.e.}, the ratio between the two types of charges. Since negative
values of $q$ lead to a divergent potential energy and taking into
account the symmetry $q \leftrightarrow 1/q$ we can restrict ourselves
to the range $0 \le q \le 1$; thus we assume that charge $Q_1$ is stronger than
charge $Q_2$. Note that we recover the classical Wigner
problem for $q=1$.

To compensate for the charges, we introduce a uniform, neutralizing
background on the plane, specified by a charge density $\sigma$, which
is given by 

\begin{equation}
\sigma = - \rho_1 Q_1 - \rho_2 Q_2 = -Q_1 \frac{1+q}{2}\rho .
\label{eq:surface_density}
\end{equation}

\section{Method}
\label{sec:method}

The present contribution is dedicated to a complete identification of
the ground state configurations of an equimolar mixture of point
charges, {\it i.e.}, the ordered equilibrium structures at vanishing
temperature. Following the basic laws of thermodynamics, the particles
will arrange under these conditions in an effort to minimize the
corresponding thermodynamic potential. For our system ({\it i.e.}, fixed
particle number $N$ and density $\rho$), we have to minimize the
potential energy, which reduces at vanishing temperature to the lattice
sum of the ordered particle configuration.

Among the numerous optimization schemes available in literature, we
have opted for an optimization algorithm that is based on ideas of
EAs \cite{ref:holland_1975,
  ref:gottwald_2005}. Our choice is motivated by the fact that this
strategy has turned out to be highly successful in related problems
for a wide variety of soft matter systems, including in particular
problems in two-dimensional geometries \cite{ref:fornleitner_2008,
  ref:fornleitner_2008_a, ref:fornleitner_2009, ref:doppelbauer_2010,
  ref:antlanger_2011}. For a comprehensive presentation of this
optimization algorithm and of the related computational and numerical
details, we refer the reader to \cite{ref:gottwald_2005,
  ref:doppelbauer_2012}.

The quantity that has to be minimized is the lattice sum of an ordered
particle configuration. Taking into account the long-range character
of the interactions (\ref{eq:coulomb}), this quantity can most
conveniently be calculated via Ewald sums \cite{ref:mazars_2011}. For
the separation of $r$- and $k$-space contributions, we have used the
cutoff values $r_c = 15 /\sqrt{\rho}$ and $k_c = 10 \sqrt{\rho}$,
respectively; for the Ewald summation parameter we use $\alpha =
0.3$. This set of numerical parameters guarantees a relative accuracy
of $10^{-5}$ for the evaluation of the internal energy.

In an effort to specify the ordered structures, we introduce for
convenience a $q$-dependent reference energy. For the one component
system ({\it i.e.}, $q = 1$) the ground state energy (per particle) of
point charges, arranged on a neutralizing plate in a hexagonal lattice,
is given by ($Q = Q_1 = Q_2$)

\begin{equation}
E_0 (q = 1) = - C_{\rm M} \sqrt{\rho} Q^2 ,
\label{eq:ref_energy_0}
\end{equation}
$ C_{\rm M} = 1.960515789$ being the Madelung constant of this
particular particle arrangement.

We extend this expression continuously to $q \le 1$ along the following
lines: we imagine the system to be split up into two infinitely large
regions, labeled $\gamma = 1$ and $\gamma = 2$, each of them hosting
exclusively the respective charges, $Q_1$ and $Q_2$, and each of them
being locally charge neutral. The two regions share a common border. Introducing local number
densities, $\rho^{(\gamma)}_i$, for species $i$ in region $\gamma$ ($i
= 1, 2$ and $\gamma = 1, 2$), we arrive at the following relations:

\begin{eqnarray} \nonumber
\rho^{(1)}_1 Q_1 + \sigma = 0 ~~~~ \rho^{(1)}_2 = 0 & & {\rm ~in~region~1} \\ \nonumber
\rho^{(2)}_1 = 0 ~~~~ \rho^{(2)}_2 Q_2 + \sigma = 0 & & {\rm ~in~region~2} .
\end{eqnarray}
Together with (\ref{eq:surface_density}), we obtain

\begin{equation}
\rho^{(1)}_1 = \frac{1+q}{2}\rho ~~~~ {\rm in~region~1} ~~~~~~~
\rho^{(2)}_2 = \frac{1+q}{2q}\rho ~~~~ {\rm in~region~2} .
\end{equation}

With these values for the local number densities and assuming that the
charges will form hexagonal lattices in the respective regions, we
obtain -- with the help of equation (\ref{eq:ref_energy_0}) -- the
total energy per particle for this system

\begin{eqnarray} \nonumber
E_0 \left( q \right) & = & 
- C_{\rm M}
\left( \frac{1}{2} \sqrt{\rho_1^{(1)}} Q_1^2 + 
\frac{1}{2}\sqrt{\rho_2^{(2)}} Q_2^2 \right) \\
& = & - C_{\rm M} \sqrt{\rho} Q^2 \sqrt{\frac{1+q}{2}} \frac{1+q^{3/2}}{2} .
\label{eq:reference_energy}
\end{eqnarray}

In an effort to characterize the emerging ground state configurations,
we have evaluated the two-dimensional orientational bond order
parameters, $\Psi_4$ and $\Psi_6$ \cite{ref:strandburg_1988}, defined via

\begin{equation}
\Psi_n = \left| \frac{1}{N_i} \sum_{j=1}^{N_i} \exp[\imath n \Theta_{j}] \right| ~~ n = 4, 6
.
\label{bond_order}
\end{equation}
$N_i$ is the number of nearest neighbours of a tagged particle with index $i$ and
$\Theta_{j}$ is the angle of the vector connecting this particle with
particle $j$ with respect to an arbitrary, but fixed orientation.

\section{Results}
\label{sec:results}

In an effort to identify the complete set of ordered ground state
configurations of our system specified in Section \ref{sec:model}, we
have performed extensive EA-runs, taking into account up to 20
particles per species and per unit cell. For a given state point, up
to 5,000 individuals were created. Calculations have been performed on
a discrete $q$-grid with a spacing of $\Delta q = 0.01$; thus in this
contribution $\Delta q$ defines the accuracy in the location of the
boundaries between ground states. The respective minimum energy
configurations were retained as the ground state particle
arrangements.

\begin{figure}[htb]
\vspace{-2ex}%
\centerline{\includegraphics[width=0.75\textwidth]{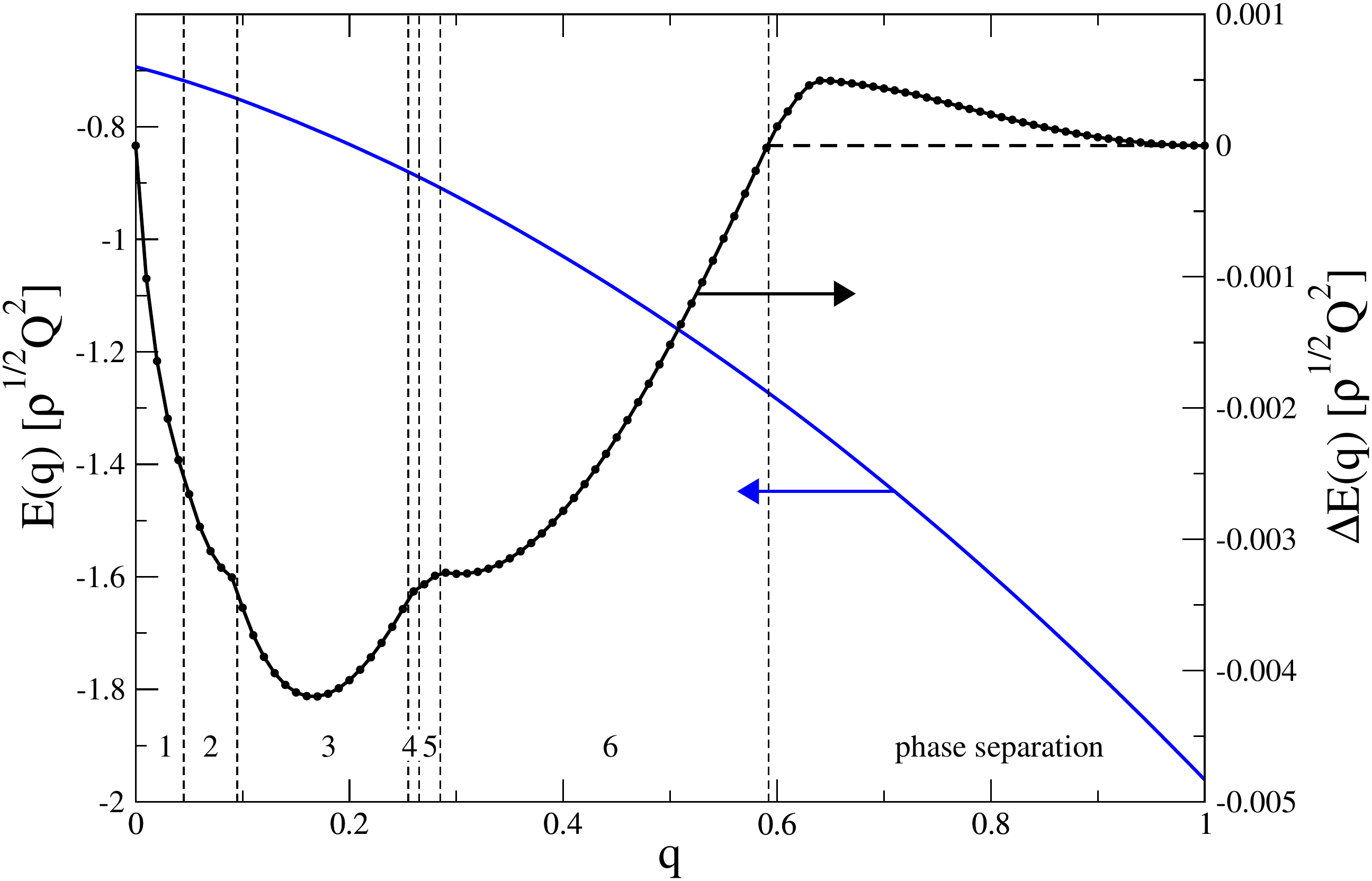}}
\caption{Energy per particle of the ground state configuration,
  $E(q)$ (blue line), and energy difference (per particle), $\Delta
  E(q) = [E(q) - E_0(q)]$ (black line), with respect to the reference
  energy, $E_0(q)$, as defined in equation (\ref{eq:reference_energy})
  as functions of $q$. Vertical broken lines indicate the limits of
  stability of the six identified non-trivial ground states. The
  horizontal broken line marks the $q$-values where the phase separated system is the energetically most favourable one.}
\label{fig:energy}
\end{figure}

In Figure \ref{fig:energy} we display the energy (per particle), $E(q)$, of these ground state configurations and the energy difference,
$\Delta E(q) = [E(q) - E_0(q)]$, with respect to the reference energy,
$E_0(q)$, as defined in equation (\ref{eq:reference_energy}). Note
that over the entire $q$-range $\Delta E(q)$ is very small, {\it
  i.e.}, less than $5 \times 10^{-3}$. The fact that the differences
between competing structures are so small is a fingerprint of the
long-range nature of the Coulomb interaction.

$E(q)$ decays with increasing $q$: it connects the limiting value at $q =
0$ [$E(q = 0) = E_0(q=0) = - C_{\rm M} \sqrt{\rho} Q^2 /2 \sqrt{2}$], obtained for
a pure system of charges $Q_1$ and a number density $\rho_1 = \rho/2$ with the other reference state at $q = 1$, where the charges
are indistinguishable ({\it i.e.}, $Q_1 = Q_2 = Q$) and thus $E(q = 1)
= E_0(q = 1) = - C_{\rm M} \sqrt{\rho} Q^2$. In the intermediate $q$-range, the
curve seems -- at first sight -- to be a smooth, monotonous
function. The subtle details, which reflect the structural changes of
the system as $q$ varies, become visible only if we subtract from
$E(q)$ the reference energy, $E_0(q)$, {\it i.e.}, $\Delta E(q) = E(q)
- E_0(q)$. This function is now non-monotonous and shows kinks for particular $q$-values which
can be associated with the structural changes. For convenience, the
vertical broken lines in Figure \ref{fig:energy} indicate the limits
of stability of the six non-trivial identified ground states. The
fact that these kinks are sometimes more or less pronounced is related
to three issues: (i) the limited accuracy of our energy evaluation
(see discussion above), (ii) the finite grid-size $\Delta q$
underlying our investigations, and (iii) the intersection angle
between the $E(q)$-curves of two neighbouring ground state
structures. For $0.59 \lesssim q <1$ the energy of the (phase
separated) reference state, $E_0(q)$, attains values that are smaller
than the energy of the respective ground states identified in our EA
search, indicating that the system undergoes a phase separation. This phenomenon, {\it i.e.}, the formation of two {\it infinitely}
large regions populated by only one species of charges representing the coexisting phases, cannot be grasped
with our EA-based optimization tool, since it relies on a {\it finite}
number of charges per unit cell.

\begin{figure}[htb]
\begin{center}
\includegraphics[width=0.4\textwidth]{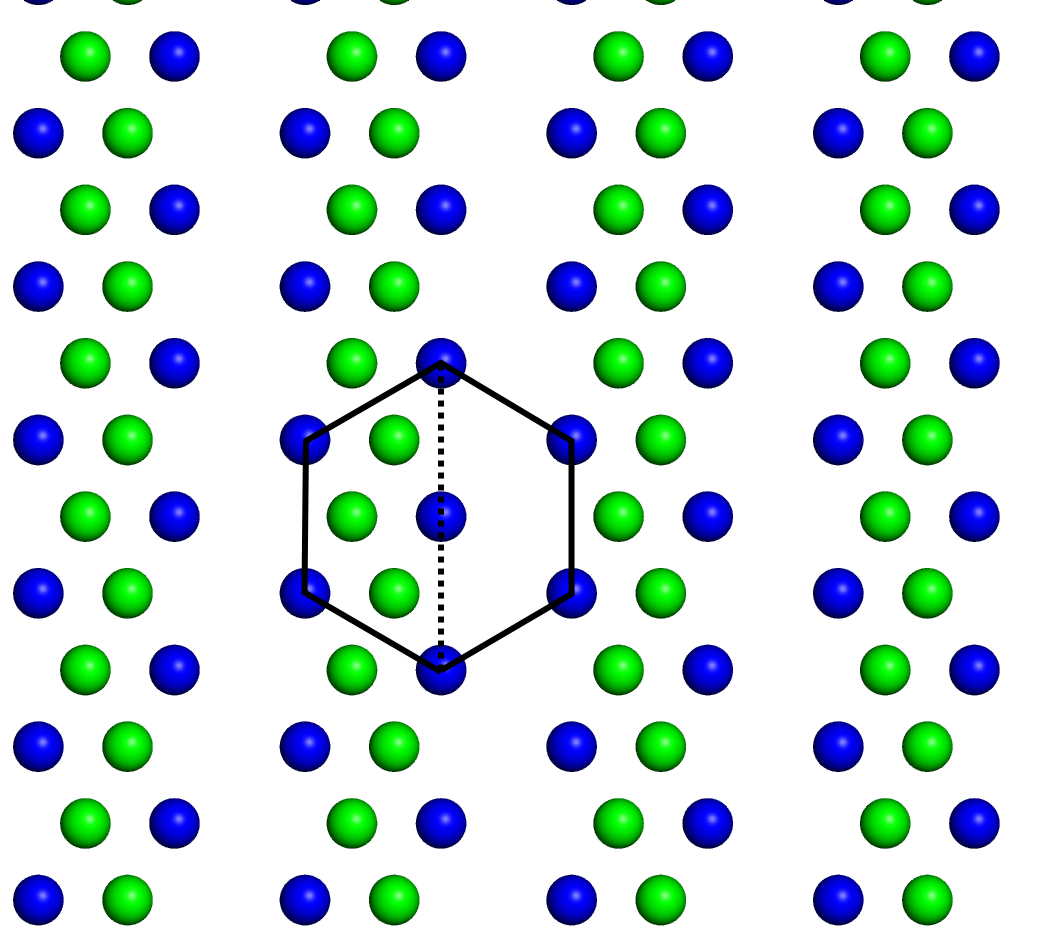}
\hspace{0.05 \textwidth}
\includegraphics[width=0.4\textwidth]{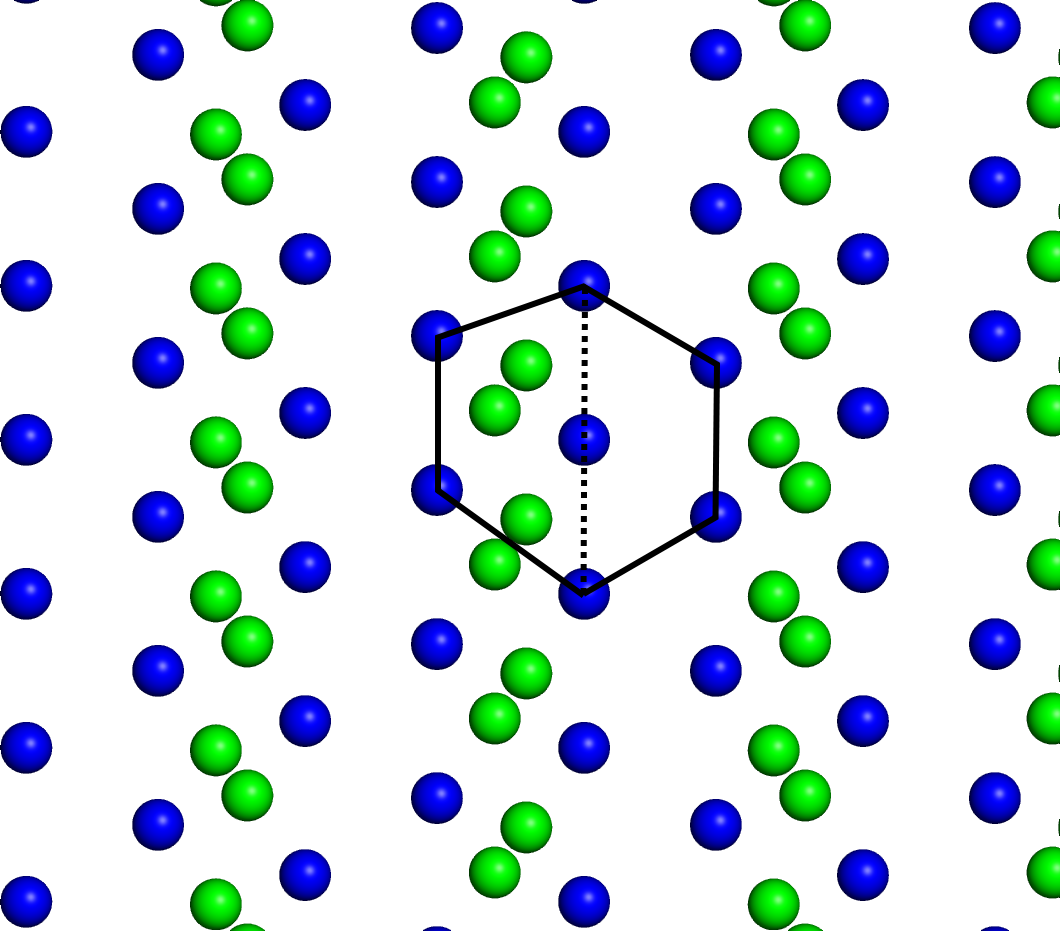}
\end{center}
\caption{Left panel: structure 1 (identified for $q = 0.02$), right
  panel: structure 2 (identified for $q = 0.05$). Blue: charges $Q_1$,
  green: charges $Q_2$. Lines highlight the hexagonal units discussed
  in the text. The dotted lines mark the central axes of the hexagonal units (cf.\ text).}
\label{fig:structure_1_2}
\end{figure}

\begin{figure}[htb]
\begin{center}
\includegraphics[width=0.4\textwidth]{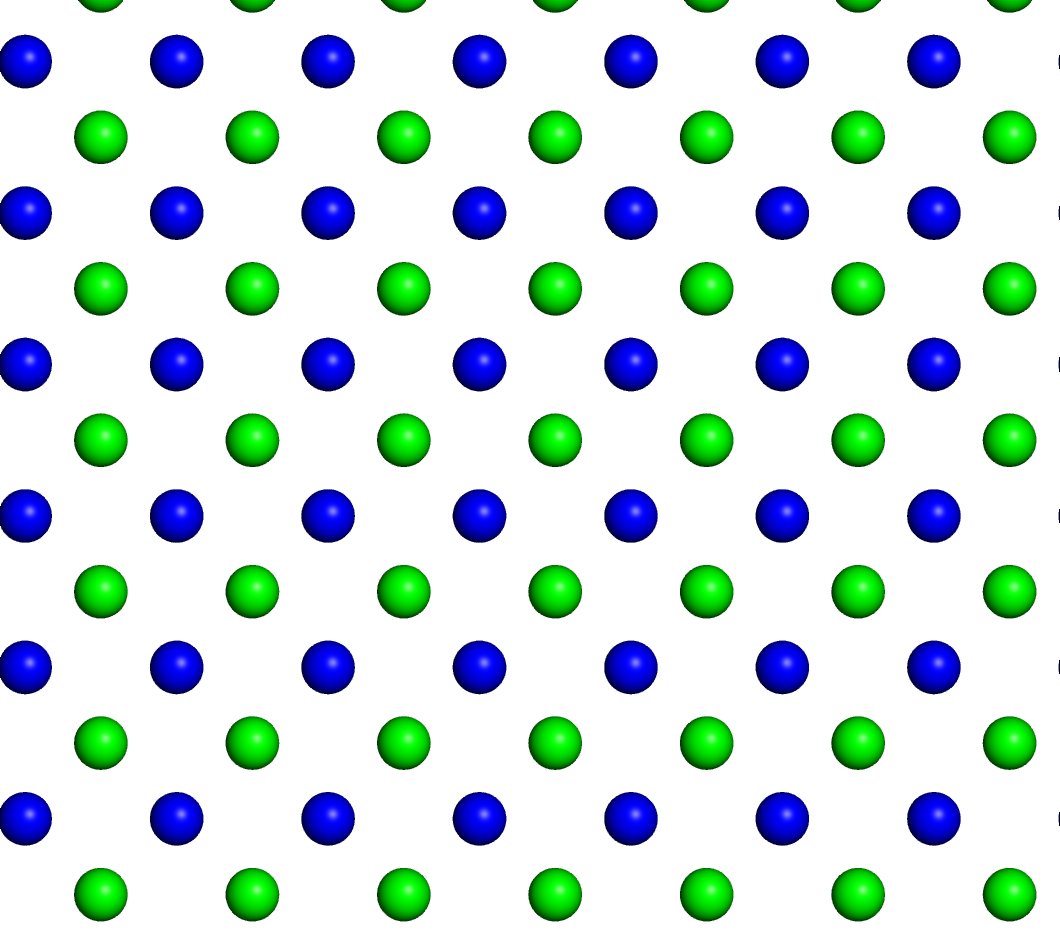}
\end{center}
\caption{Structure 3 (identified for $q = 0.10$). Blue: charges $Q_1$,
  green: charges $Q_2$.}
\label{fig:structure_3}
\end{figure}

\begin{figure}[htb]
\begin{center}
\includegraphics[width=0.4\textwidth]{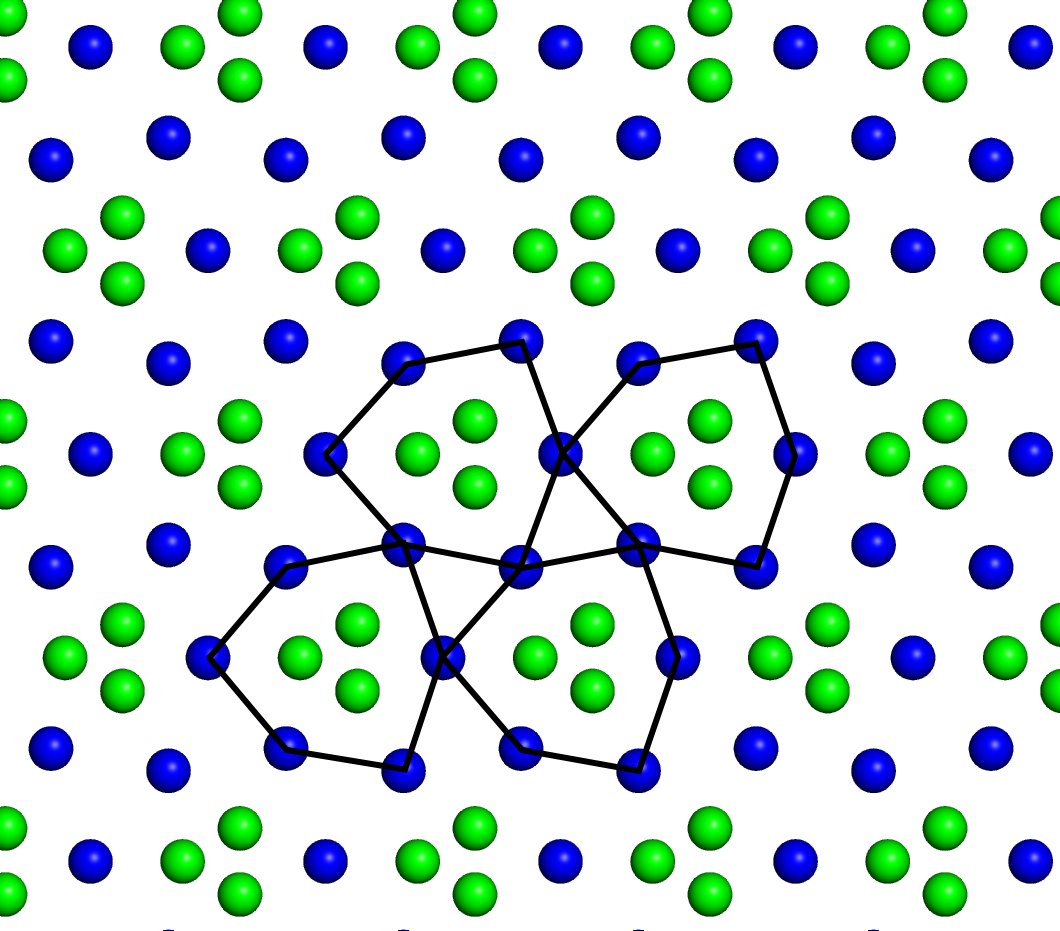}
\hspace{0.05 \textwidth}
\includegraphics[width=0.4\textwidth]{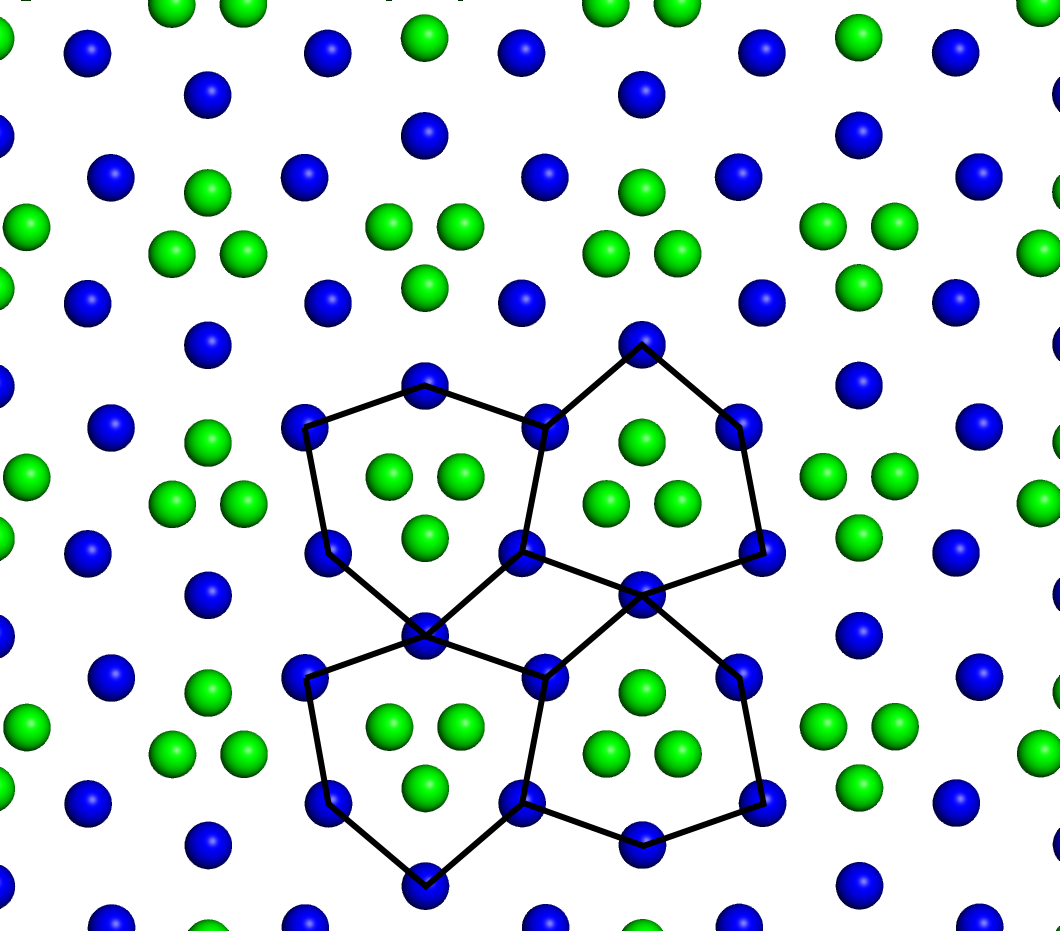}
\end{center}
\caption{Left panel: structure 4 (identified for $q = 0.26$), right
  panel: structure 5 (identified for $q = 0.27$. Blue: charges $Q_1$,
  green: charges $Q_2$. Lines highlight the hexagonal units discussed
  in the text.}
\label{fig:structure_4_5}
\end{figure}

\begin{figure}[htb]
\begin{center}
\includegraphics[width=0.4\textwidth]{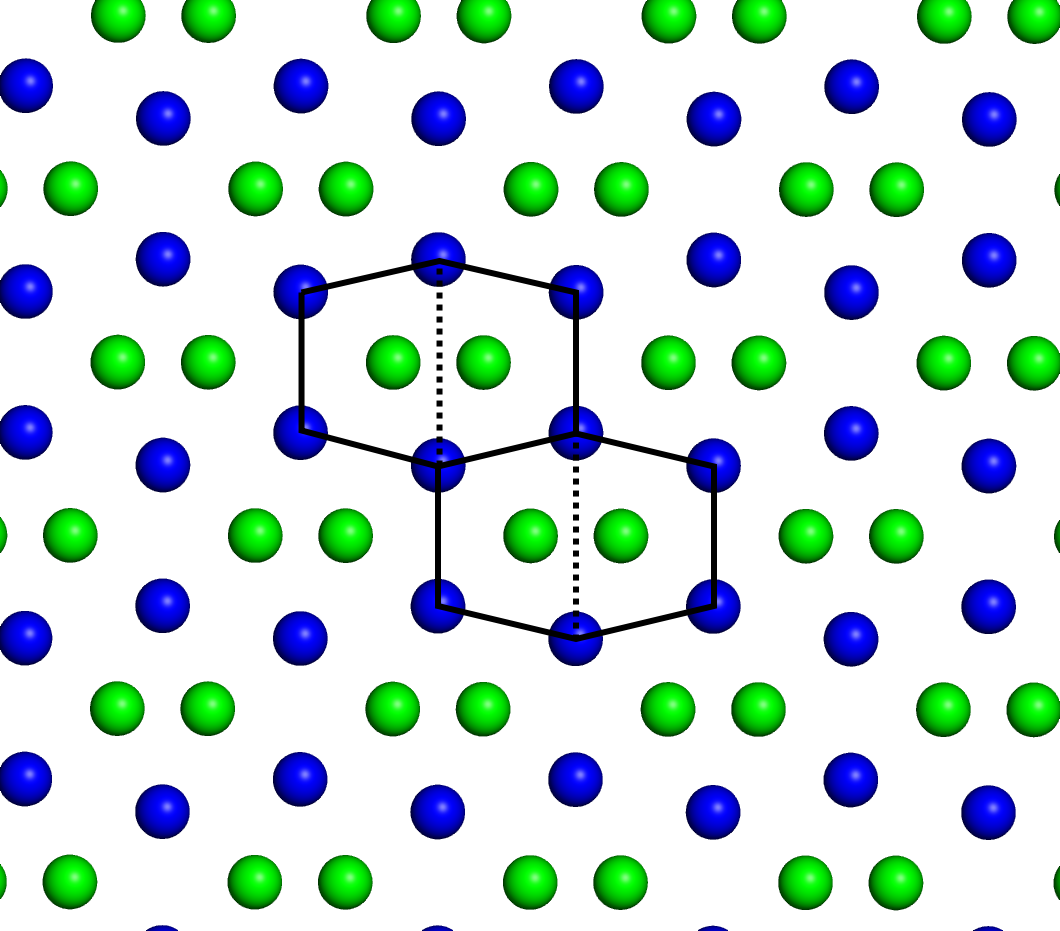}
\hspace{0.05 \textwidth}
\includegraphics[width=0.4\textwidth]{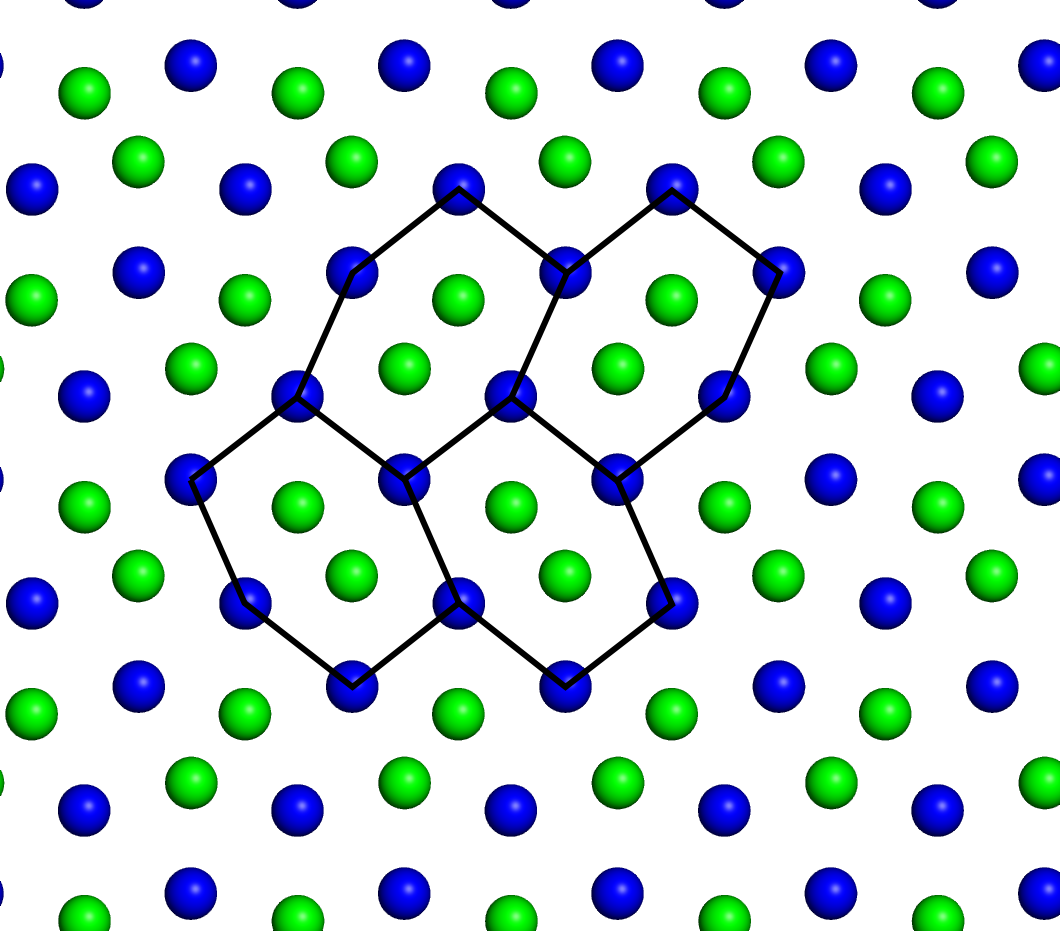}
\end{center}
\caption{Left panel: structure 6 (identified for $q = 0.30$), right
  panel: structure 6' (identified for $q = 0.35$. Blue: charges $Q_1$,
  green: charges $Q_2$. Lines highlight the hexagonal units discussed
  in the text. The dotted lines mark the central axes of the hexagonal units (cf.\ text).}
\label{fig:structure_6}
\end{figure}

The Table provides an overview over the ground state
configurations that we have identified for $0 \le q \le 1$; the
structures themselves are depicted in Figures \ref{fig:structure_1_2}
-- \ref{fig:structure_6}.

\begin{table}
\caption{Overview over the identified ground state configurations and the respective $q$-ranges. The
  structures themselves are depicted in Figures
  \ref{fig:structure_1_2} -- \ref{fig:structure_6}.}
\label{tab:table}
\begin{center}
\begin{tabular}{cll} 
$q$-range & ground state structure \\
\hline
0.00         & hexagonal lattice formed by charges $Q_1$  \\
$0.00 < q \lesssim 0.04$ & Structure 1 \\
$0.05 \lesssim q \lesssim 0.09$ & Structure 2 \\
$0.10 \lesssim q \lesssim 0.25$ & Structure 3 \\
$0.26 \simeq q $ & Structure 4 \\
$0.27 \lesssim q \lesssim 0.28$ & Structure 5 \\
$0.29 \lesssim q \lesssim 0.59$ & Structure 6 \\
$0.60 \lesssim q < 1$ & phase separation \\
1.00                  & hexagonal lattice formed by the \\
                      & (indistinguishable) charges $Q_1$ and $Q_2$ \\
\hline
\end{tabular}
\end{center}
\end{table}

For \boldmath$q = 0$\unboldmath, charges with non-vanishing nominal values, $Q_1$,
arrange -- as expected -- in a hexagonal lattice (not depicted) with
number density $\rho_1 = \rho/2$; the other, chargeless particles do not
interact with any particle species and thus occupy arbitrary
positions.

For \boldmath$0 < q \lesssim 0.04$\unboldmath, charges $Q_1$ form a hexagonal
lattice which is to a high degree regular (Structure 1, depicted in
the left panel of Figure \ref{fig:structure_1_2}): the order parameter
$\Psi_6$ varies between $\Psi_6 (q = 0.01) = 0.99972$ and $\Psi_6(q =
0.04) = 0.99577$. The deviation from its ideal value, $\Psi_6 = 1$,
stems from a slight distortion of these hexagons (highlighted in the corresponding panel): a central 'axis' (dotted line in the corresponding panel),
connecting the 'upper' and 'lower' vertices of the hexagon, is formed
by two line-segments of equal length; the distances of the vertices
left (right) to this central axis from the center of the hexagon
differ in their length by less than +4\% (-3 \%). Thus the left half
of the hexagon is slightly larger than its counterpart located to the
right of the central axis. This enlarged space is imposed by the fact
that this area hosts charges $Q_2$, which form a zig-zag pattern
within the ground state configuration, oriented parallel to the central
axis of the hexagon; within this line, charges $Q_2$ are
equidistant. The other part of the hexagon, located right to its
central axis remains empty.

In the adjacent $q$-range, \boldmath$0.05 \lesssim q \lesssim 0.09$\unboldmath,
charges $Q_1$ maintain their hexagonal arrangements (Structure 2,
displayed in the right panel of Figure \ref{fig:structure_1_2}). As
compared to Structure 1, the distortion of the hexagons (highlighted in the corresponding panel) is now
considerably more pronounced: $\Psi_6(q = 0.05) = 0.90598$ while
$\Psi_6(q = 0.09) = 0.83643$. The 'upper' and the 'lower' vertices of
the hexagon are still connected via a central axis (dotted line in the corresponding panel), consisting of two
line segments of equal length; however, the distances of the vertices
left to this central axis from the center of the hexagon are now by up
to 21 \% longer, while the corresponding distances of the vertices on
the opposite side of the axis are less than 5 \% shorter. This
conceivable asymmetry (see the highlighted hexagon in the corresponding panel) is imposed by the increased nominal value of
charge $Q_2$: the zig-zag arrangement of particles, observed for this
species of charges in Structure 1 has been replaced by a parallel
arrangements of pairs of particles which are aligned in the direction
of the central axis of the hexagon; while the intra-pair distance is very
short, the inter-pair distance is quite large.

In the relatively wide range of \boldmath$0.10 \lesssim q \lesssim 0.25$\unboldmath~the two species of charges arrange in two intertwining, commensurate
square lattices (Structure 3, cf. Figure \ref{fig:structure_3}). It
has to be emphasized that both sublattices remain perfect over the
entire $q$-range of stability, {\it i.e.,} $\Psi_4 = 1$ for $0.10
\lesssim q \lesssim 0.25$. The structural stability of this particular
ground state is also reflected by the fact that Structure 3 is
characterized by the highest energy gain compared to the energy value of
the reference structure, $E_0(q)$ (see Figure \ref{fig:energy}).

Around the value \boldmath$q \simeq 0.26$\unboldmath, charges $Q_1$ form hexagonal
structural units which are in their shape reminiscent of gems or
diamonds.  These six-particle rings form in a head-to-tail arrangement
{\it parallel} lanes: adjacent six-particle rings of neighbouring
lanes share vertices, while the remaining edges form equilateral
triangles. This ground state is referred to as Structure 4; it is
depicted in the left panel of Figure \ref{fig:structure_4_5} and the six-particle rings are highlighted. Each of
the six-particle units hosts in its center an essentially equilateral triangle of charges $Q_2$. The positions of the six
surrounding charges $Q_1$ are imposed by the condition that the
smallest distance of these charges from any of the three inner charges
($Q_2$) has the same value; this requirement induces the particular
shape of the six-particle rings. The triangular arrangements that fill
up the interstitial space are not populated by charges $Q_2$.

The ground state identified for \boldmath$0.27 \lesssim q \lesssim 0.28$\unboldmath~(denoted as Structure 5 and depicted in the right panel of Figure
\ref{fig:structure_4_5}) differs only in one feature from Structure 4:
the lanes, formed by the head-to-tail arrangements of the six-particle
rings (highlighted in the corresponding panel) are now {\it antiparallel}. In this configuration the neighbouring rings of adjacent lanes share edges, which leads now to the formation
of rhombic four-particle arrangements which are again void of $Q_2$
charges.

Finally, at $q \simeq 0.29$ Structure 6 emerges and remains the
ground state over the relatively large interval \boldmath$0.29 \lesssim q
\lesssim 0.59$\unboldmath~(see left panel of Figure \ref{fig:structure_6}). Its
basic unit is an elongated hexagon (highlighted in the corresponding panel): aligned in parallel and sharing
edges with neighbouring tiles, they completely cover the
two-dimensional space. The direction perpendicular to the longest
elongation of this hexagon is considered for the following discussion
as the central axis (dotted in the corresponding panel). For Structure 6 this axis is also the symmetry axis of the hexagon. The four edges originating from the central
axis have the same lengths, say $l_1$; similarly, the remaining two
edges of the hexagon (oriented parallel to the central axis) assume
another, equal value, say $l_2$. Each of these hexagons hosts a pair
of charges $Q_2$, located on a line perpendicular to the central axis and
separated by a distance, which decreases as $q$ is
increased. By increasing the charge ratio $q$,
$l_1$ decreases from $1.297\sqrt{\rho}$ (at $q = 0.29$) to $1.269\sqrt{\rho}$ (at $q = 0.59$),
while $l_2$ increases from $1.29\sqrt{\rho}$ (at $q= 0.29$) to $1.324\sqrt{\rho}$ (at $q =
0.59$). At the cross-over, {\it i.e.},  $l_1 \simeq l_2$ (observed for
$q \simeq 0.36$, our optimization tool identifies a closely related,
energetically degenerate structure, denoted as Structure 6' and
depicted in the right panel of Figure \ref{fig:structure_6}: now that
all edges of the basic hexagon are equal, these units are no longer
forced to align in parallel, but are able to chose an alternative,
non-parallel arrangement: imposed by the internal angle between edges
of the basic hexagon, these units arrange in a grain-like
super-structure.

For \boldmath$0.59 \lesssim q < 1.00$\unboldmath, we find that $E_0(q) < E(q)$ (see Figure
\ref{fig:energy}), indicating that the phase separated
particle configuration is energetically more stable than any other
ordered structure identified by our optimization tool. Due to the
limitation in the number of particles per unit cell, the EA-based
search for ground state configurations proposes -- depending on the
number of particles per cell -- configurations with increasing
complexity, all of them being characterized by an energy value $E(q)$
that is larger than the corresponding value $E_0(q)$. Thus the demixed
state, formed by two separate hexagonal, ordered regions and each of
them being populated by one species of charge, is the ground
state in this $q$-range.

Finally, for \boldmath$q=1$\unboldmath, we recover the one-component hexagonal
monolayer (not displayed).

\section{Conclusions}
\label{sec:conclusions}

In this contribution we have investigated the ground state
  configurations of an equimolar, binary mixture of classical charged
  particles (with nominal charges $Q_1$ and $Q_2$), that self-assemble
  on a neutralizing plane. Our investigations are based on reliable
  Ewald summation techniques which allow an efficient evaluation of
  the ground state energies (= lattice sums). With the help of
  reliable optimization tools, which are based on ideas of evolutionary
  algorithms, we are able to identify the ordered ground states of the
  system: by searching essentially among all possible two-dimensional
  lattices, this algorithm identifies for a given charge ratio $q =
  Q_2 / Q_1$ the energetically most favourable particle
  arrangement. Apart from the expected, trivial hexagonal lattices for
  $q = 0$ and $q = 1$, we could identify for $0 < q \lesssim 0.59$ in total
  six ground state configurations.

Quite unexpectedly, these particle arrangements show a remarkable
structural complexity which is the result of the energetic competition
between the charge-charge interactions. Throughout, a pronounced
impact of the weaker charges, $Q_2$, on the sublattices formed by the
stronger charges, $Q_1$, could be observed: this holds even if the
corresponding $q$-values are rather small ({\it i.e.}, $q \simeq
0.05$). Except for a purely square particle arrangement (which is
stable over a remarkably large $q$-range and which shows the highest
energy gain among all ground states with respect to a suitably defined
reference state -- see also the discussion about energies below), the
ground states can be described on the basis of asymmetric, sometimes
strongly distorted six-particle arrangements formed by charges $Q_1$,
which host in their interior simple two- or three-particle
configurations of charges $Q_2$. A deeper insight into the mechanisms
that govern the formation of ground states is gained by introducing a
phase separated reference state, where the two species of charges
populate in hexagonal arrangements spatially separated areas.
Comparing the energies of our ground state configurations, $E(q)$,
with the energy of this reference state, $E_0(q)$, we find that these
two functions differ by less than $5 \times 10^{-3}$, a fact that
represents a characteristic fingerprint of the long-range Coulomb
interactions. An analysis of this energy difference as a function of
$q$ reveals that transitions from one ground state to an adjacent one
become visible as (more or less pronounced) kinks in this
function. Based on these considerations we could show that for $0.59
\lesssim q < 1$ the energetically most favourable particle arrangement
is the (ideal) phase separated state.

\section*{Acknowledgments}

This work was financially supported by the Austrian Research Fund
(FWF) under Proj.\ No.\ P23910-N16. The authors gratefully acknowledge
discussions with Martial Mazars (Paris-Orsay), Ladislav {\v S}amaj
(Bratislava), and Emmanuel Trizac (Paris-Orsay).

%% Type in your references using {thebibliography} environment 
%% or create them from your bibtex database using cmpj.bst style (experimental).

%\bibliographystyle{cmpj}
%\bibliography{mybibdb}

\end{document}